\documentclass[twocolumn,twoside,slac,nofootinbib]{revtex4}
\usepackage{graphicx}
\usepackage{fancyhdr}
\begin{document}

\title{Magnetic monopoles and cosmic inflation}
\author{Alfred Scharff Goldhaber}
\affiliation{C.N.Yang Institute for Theoretical Physics, SUNY Stony Brook, NY 11794-3840 USA}

\begin{abstract}
It is possible that the expansion of the universe began with an inflationary phase, in which the inflaton driving the process also was a Higgs field capable of stabilizing magnetic monopoles in a grand-unified gauge theory.  If so, then the smallness of intensity fluctuations observed in the cosmic microwave background radiation implies that the self-coupling of the inflaton-Higgs field was exceedingly weak.
It is argued here that the resulting broad, flat maximum in the Higgs potential makes the presence or absence of a topological zero in the field insignificant for inflation.  There may be monopoles present in the universe, but the universe itself is not in the inflating core of a giant magnetic monopole.

\end{abstract}

\maketitle

\section {\bf Introduction -- Topology as a possible alternate
source of inflation}

Two phenomena that should occur on totally different length scales nevertheless may be connected quite closely.  Magnetic monopoles are classical-field solutions of grand-unified gauge theories, expected to survive quantum corrections and fluctuations because the poles are associated with long-range fields that would require infinite action to destroy.   't Hooft \cite {'t H} and Polyakov  \cite {polya} independently discovered the first such solution in SO(3) gauge theory with the apparent symmetry reduced to U(1) by the Higgs mechanism \cite{Higgs:1964ia}.  Kibble \cite{kib} observed that in a universe without gravity and with the vacuum value of the Higgs field initially very small, quantum fluctuations of that field would produce structures with the topology of global monopoles, which could be deconfined by acquisition of magnetic monopole gauge fields.  Preskill \cite{Preskill:1979zi} realized that this possible monopole production could create a crisis for cosmology, implying far more monopoles than observational limits allow.  Because the expected energy scale of grand unification is quite  high, the geometrical size of a monopole core must be quite small.

At the other extreme of length comes the visible universe.  Guth \cite{guth} introduced a concept which may underlie the origin of the universe, cosmic inflation.  The essence is that a dynamical scalar field, the {\it inflaton}, initially may have in some region an expectation value which leaves the energy density of the field far from its minimum at zero energy density.  This implies a dynamical (therefore variable) coefficient for Einstein's cosmological term, and hence exponential expansion, or inflation, of the region, coming to an end when the inflaton field reaches  the value which minimizes the energy density.

 At first sight it might seem that such a model would not be amenable to scientific investigation, because science exploits insights about structure to predict future observations.   Nevertheless, as abundantly illustrated in a field like geology, it is possible to predict future observations related to events long ago, because we keep developing new methods to observe consequences of those events.  In the case of inflationary cosmology, exactly such developments are occurring, in particular through improved sensitivity to angular fluctuations in the cosmic microwave background radiation, most recently with WMAP   \cite{Spergel:2003cb}.\footnote{The scale-invariant spectrum of angular fluctuations  appearing in many inflationary models was suggested even earlier on grounds of simplicity by  Harrison and by Zeldovich  \cite{harzel}.  Thus this phenomenon may be a consequence of inflation, but its confirmation would not necessarily single out inflationary cosmology as the only possible origin of such a spectrum.  To accomplish that would require more detail, possibly forthcoming in future observations.}
 
 How is all this connected to magnetic monopoles? Guth \cite{guth} observed that one
consequence of inflation is the disappearance of the monopole
problem:  Inflation sweeps apart all the monopoles implied by the Kibble mechanism, 
generating our entire visible universe from a tiny
preinflation region, which might not contain even one pole.  However, the possible connection might be even stronger, because the Higgs field stabilizing a monopole in a  grand unified model is a candidate to play the role of the inflaton.   Let us return to this possibility shortly.

The original Guth proposal had defects, some of which were addressed by `new inflation', independently proposed by Albrecht and Steinhardt \cite{Albrecht:1982wi} and Linde \cite{Linde:1981mu}.   The next (and still viable) stage was the `chaotic eternal inflation' of Linde \cite{lin1}, and Steinhardt \cite{steinh1}, and Vilenkin \cite{vil1}, where, the process of inflation never comes to a complete end. The transition from false to true vacuum
occurs in some parts of the Universe, while the rest remains in the
false vacuum state. In this way the global structure of the
Universe is the following: thermalized phase surrounding islands of
inflating space, or islands of thermalized phase surrounded by
false vacuum, depending on the rate of formation of bubbles of the
new phase. This eternal process is due to quantum fluctuations of the inflaton
field, which in localized regions can inhibit the field from
rolling down towards the minimum of its effective potential.\footnote{This description is drawn from the elegant formulation in  \cite{     lin2}.}

In 1994 Linde  \cite{lin2} and  Vilenkin \cite{vil2}
independently proposed an alternative way to
generate eternal inflation: In the center of a topological defect
the scalar field is zero, and hence at a local maximum of its
effective potential. This looks like a suitable condition for the
onset of  inflation.
 Because the topological defect is indestructible it appears that,
once started, inflation of the core will never end. Further,
with this scheme there should be no need for fine-tuning of the initial
state.  

Numerical investigation of this proposal was performed in
\cite{sak2} for a global monopole which starts out as a static
solution of
 classical field equations in the absence of gravity and
Higgs-field self-coupling and in \cite{sak1} for an 't
Hooft-Polyakov magnetic monopole \cite {'t H,polya} with the same
assumptions for initial conditions.  Such a calculation looks as if it might be
relevant for a case in which the vacuum expectation value (vev) that
minimizes the Higgs potential, initially small, slowly increases at all
points in space to
a value at or beyond the Planck scale, where the gravitational interactions
become comparable with gauge and scalar field contributions:  The
monopole's gauge and
scalar fields rescale gradually, always solving the static Higgs-Yang-Mills-Einstein
equations with the current vev, until a point is reached at which this structure becomes unstable.

 If one kept using the original vanishing Higgs self-coupling, then the instability would be towards collapse, thanks to the positive gravitational mass density of the gauge field, while if the Higgs self-coupling were increased, the Higgs field in the exterior region would approach its asymptotic vacuum value exponentially with increasing $r$, rather than simply as $1/r$.  This implies that  the starting points for the numerical calculations described above no longer should be appropriate for large self-coupling.  

Given observational constraints on the self-coupling, even a correct numerical calculation
with large self-coupling would not describe a
realistic scenario.  However, if it did yield inflation, then this would
imply  topological
inflation in a theory with these parameters under almost any choice of initial
conditions.  On the other hand, for the extremely small self-coupling implied by observation, the main message of this paper is that nontrivial topology of the Higgs field is unimportant for inflation.  Some of the considerations here were discussed briefly in earlier work, but with a more optimistic conclusion about the importance of topological inflation.  If the present analysis is right, then the earlier work did not pursue these issues quite far enough.

The concept of topological inflation certainly is correct for at
least some examples.  It is quite straightforward in the case of a
domain wall for large enough vev of an inflaton field with nonzero self-coupling 
\cite{lin2,vil2,j1}.  Still, among all the possible cases, inflation of
a magnetic monopole core holds special appeal.    The
attraction of such a scheme would only be increased by the economical notion
that the Higgs field which stabilizes monopoles could also be the
inflaton field.\footnote {This notion did not receive much consideration before the work of Linde and Vilenkin, probably because of the rather large expected scale for the Higgs self-coupling, suggested by perturbative quantum field theory.  Phenomenology of the observed cosmic density variations suggests a much smaller coupling if inflation were the immediate precursor of our universe.  These points are elaborated here a little later.}

Thus, inflation disconnects us from monopoles, but it remains possible that monopoles are connected to, and even generate, inflation.  The main thesis of the present work, based on strong evidence from cosmic microwave background (CMB) fluctuations that there was very weak inflaton self-coupling during  inflation preceding our present expanding universe, is that such inflation in no way was seeded or enhanced by the formation of zeros in an inflaton-Higgs field carrying the topological charge of a magnetic monopole.  In other words, our universe might contain one or a few magnetic poles, but it does not constitute the inflated core of a magnetic pole.  

Previous literature contains the observation that monopole topological inflation (with a rather large Higgs self-coupling) could have occurred in an earlier phase of inflation.  This would get round the problem of inconsistency between naive expectations for a rather large Higgs self-coupling and the small inflaton self-coupling required for inflation directly initiating our current universe, but at the price of making the relationship between us and the inflating monopole core exceedingly remote!
  The main issue addressed in the present work, with a negative conclusion, is whether a {\it small} self-coupling for the Higgs field would lead to monopole topological inflation, as distinguished from chaotic inflation.

\section{\bf Previous analyses}
 For the numerical calculations in \cite{sak1} Sakai considered
Yang-Mills and scalar fields minimally coupled to gravity. The
action for this system is

\bigskip

\bigskip

$$S= \int d^4x \sqrt{-g} \bigg{[}
\frac{m^2_{\rm Pl}}{16 \pi}R -
\frac{1}{4} \left(F^a_{\mu \nu}
\right)^2 $$
\begin{equation} - \frac{1}{2} \left (D_{\mu} \Phi^a
\right)^2 - V(\Phi) \bigg{]}\label{sakai}
\end{equation}
\newline where $\Phi$ is a real triplet, and $F^a_{\mu
\nu}$ is the field strength of the SU(2) gauge field
$A^a_{\mu}$:
\begin{equation} F^a_{\mu \nu}= \nabla_{\mu} A^a_{\nu} -
\nabla_{\nu} A^a_{\mu} - e
\epsilon^{abc} A^b_{\mu} A^c_{\nu},
\end{equation}
$D_{\mu}$ is the fully covariant derivative:
\begin{equation} D_{\mu} \Phi^a = \nabla_{\mu}
\Phi^a + e\epsilon^{abc}A^a_{\mu}\Phi^c,
\end{equation} ($\nabla_{\mu}$ being the spacetime covariant
derivative), and the potential of the scalar field is:

\begin{equation} V(\Phi) = \frac{1}{4} \lambda
\left( \Phi^2 - \eta^2 \right)^2.
\end{equation} \\ In \cite{sak1} the field equations were solved
numerically for different values of the parameters - the vev $\eta$
of the scalar field and the ratio $ \frac {\lambda}{e^2}$ of the
two coupling constants. The results were: Assuming a static initial
configuration that solves the (first-order)
Bogomol'nyi-Prasad-Sommerfield [BPS] equations \cite{bps}, which
hold for Higgs self-coupling $\lambda=0$, then for
$\eta$  order of the Planck mass
$m_{\rm Pl}$ and
$\lambda/e^2$ order unity inflation does occur. The
studies indicated that with
 $\eta \approx m_{\rm Pl}$ monopole inflation (that is, expansion of the core
 of an initially present mononopole)  is even possible for smaller
values of
$\lambda/e^2$, but the smallest value considered was
$\lambda/e^2=0.1$.  If the goal were to supply starting
conditions corresponding to equilibrium without gravity, then
the BPS ansatz would be questionable for
$\lambda\ne 0$ because the self-coupling suppresses the difference
$\Delta
\Phi$ between the Higgs field and the vev, forcing exponential
decay with radius
$r$, while the ansatz of course keeps $\Delta \Phi$ independent of $\lambda$,
falling only as $1/r$, and thus implies additional (spurious) contributions
to the cosmological term which drives inflation.

If the self-coupling of the scalar field arose as a perturbative
quantum effect, as in the  Coleman-Weinberg calculation \cite{cw},
one would have
$\lambda \approx e^4/4\pi$,and hence
$\frac{\lambda}{e^2}
\approx e^2/4\pi \ll 1$, because for the perturbative expansion to
make sense one has to have $e < 1$. Beyond this theoretical
argument, phenomenological considerations give a far stronger
version of the  condition. Comparison of inflation-model
predictions with astronomical observations (data on the CMB fluctuations) yield an upper limit
$\lambda\le 10^{-10}$ \cite{guth2,shafi,brand,lin3}.   At the same
time, the slow (logarithmic) running of gauge couplings with scale
assures that a grand unified magnetic monopole would be associated
with a value $e^2 \approx 0.1-1.0$.  This implies
$\frac{\lambda}{e^2}\le10^{-9}$, and is commonly taken to show
that the inflaton is not the Higgs field of a grand unified theory,
and therefore has no direct connection to the monopoles in such a
theory.  As we shall discuss later, supersymmetry, if present, could
greatly suppress the Higgs self-coupling, perhaps making it consistent with
the mentioned CMB limit.  It's worth noting that mysteriously small couplings
abound in the current scheme of physics.  Whether or not these are explained by
supersymmetry, one might take an empirical view that a small self-coupling of the
Higgs field in grand-unified gauge theory would be no more or less surprising than the
small masses of quarks and leptons, and the really small magnitude of an 
apparent cosmological term driving accelerated expansion of our universe.

We see that accepted values of the second parameter are restricted to the
range
$\frac{\lambda}{e^2}
\ll 1$.  We shall encounter below analytic evidence that rules out
monopole core inflation for small
$\lambda/e^2$.  All of this depends on choosing initial
conditions for a monopole in equilibrium without gravity, and
following the evolution when gravity is included (as was attempted (using a simplified ansatz)
in the numerical calculations mentioned above).  We shall need to
consider later a very different and arguably more appropriate choice
of initial conditions, where the expectation value of the Higgs
field everywhere starts out close to zero, but with a very small, but nonzero, Higgs
self-coupling calling for a nonzero vev.  Under these
circumstances, topological zeros must arise in the Higgs field,
and one may study the influence of their presence on the evolution.

The existence of a solution of the Yang-Mills-Higgs system coupled
to gravity (an 't Hooft-Polyakov monopole in curved background) was
shown by van Nieuwenhuizen, Wilkinson, and Perry
\cite{van}.  As in the flat-space case with nonzero $\lambda$, this
solution cannot be written in closed analytic form. Some time after
that analysis, there were several (mostly numerical)
investigations of a different aspect of
 large $\eta$: Ignoring the possibility of inflation, what happens
in a theory with both classical gravity and a Higgs-Yang-Mills
action, so that two types of monopole, a Reissner-Nordstrom
(extremal black hole) and an 't Hooft-Polyakov hedgehog, both are
possible?  In general, the naive value of the mass of the first or
the second will be greater, depending on whether $m_{\rm Pl}$ is
greater or smaller than $\eta$.  An obvious expectation is that
the more massive one (as determined by naive estimates) will be
unstable.  Instability of the Reissner-Nordstrom solution for
$\eta < m_{\rm Pl}$ was noted by Goldhaber \cite{asg}, Lee, Nair,
and Weinberg \cite{lee1}, and Breitenlohner, Forgacs, and Maison
\cite{breit}.  Instability of the 't Hooft-Polyakov monopole for
$\eta>{\cal O} (m_{\rm Pl})$ was noted by Frieman and Hill \cite
{FH},  Lee et al. \cite{lee2}, and Ortiz \cite{ortiz}, as well as
Breitenlohner et al. \cite{breit}. Even though this was not discussed explicitly in the works just cited, it is quite possible that for sufficiently
large $\lambda/e^2$ the instability is associated with monopole
inflation.

In \cite{breit} new static solutions of the spherically symmetric
Einstein-Yang-Mills [EYM]-Higgs system were found, which disappear
when gravity is decoupled. These solutions are a discrete family
of radial excitations with an increasing number of zeros of the
gauge field. In the limit $\eta \rightarrow 0$ they join smoothly
to the (again numerical) Bartnik-McKinnon solutions of the EYM system \cite{bart},
which may be viewed as gravitationally bound classical
`glueballs', and of which all but the two lowest are known to
be unstable.  Arguably these two are unstable as well (though quite possibly
metastable), because there is no obvious lower bound for the mass of
such an object. 

 Let us  concentrate for our analytic estimates of `soliton explosion'  on the 't Hooft-Polyakov
monopole, coupled to gravity. As mentioned, the  method to be used is designed for solutions having a zero-gravity limit. The same is
true also for the numerical work \cite{sak1}. The solutions
 of
\cite{breit}, which are  intrinsically nonperturbative in the
gravitational coupling, in principle could give a new
result for inflation conditions, even though they do not carry net topological
charge.  However, this seems unlikely because the gauge-field energy, more than the
Higgs-field self-coupling energy, evidently increases for these solutions
-- that generates positive gravitational mass, and hence
gravitational binding, precisely the opposite phenomenon to the inflation
we are seeking.  Further, as already mentioned, these excitations may 
be unstable against ordinary disintegration, as opposed to explosive inflation.

There is an interesting further consideration here: Much discussion
in the literature points to the necessity of introducing some
physics besides classical inflation to describe the past,
including the case of a universe generated from a bubble of false
vacuum \cite{farhi}, and in particular, as discussed by
Borde, Trodden and Vachaspati, a universe
generated by monopole inflation \cite {borde}. Indeed, the inevitability
of a past singularity for any region of a classical inflating spacetime
was indicated by Borde and Vilenkin \cite{bv}. A recent theorem
based on very weak assumptions appears to make this constraint
inescapable \cite{borde'}. The required additional physics
might be supplied by a quantum fluctuation at or later than the point in
the past where purely classical backward evolution would have become
incomplete
\cite {vil3}.

This convincing case for something in inflation beyond classical
field theory suggests a thought experiment for
numerical calculations of the type carried out by Sakai \cite{sak1}:
After verifying that inflation occurs starting from an initial condition
for a monopole coupled to gravity, one then could run the equations
backwards, and look for the past singularity.
Clearly the time symmetry of the system is such
that if at $t=0$ one had a static solution, its past history would simply
be the reflection of its future evolution.  If so, indeed in this model
inflation would not be eternal from the past -- instead there would be
inflation looking backwards, equivalent to past contraction terminating at
the equilibrium point.  This also
would violate singularity theorems about future singularities of
collapsing systems.  The obvious resolution is that there must be a
quantum fluctuation leading to any such solution, and moreover there must
be time-dependence at every stage (no static solution at any time),  increasing the conventional matter
energy and so further raising the threshold for inflation.

 The next two sections are devoted to examining the influence of the two dimensionless parameters
in the
EYM-Higgs system,
$\lambda/e^2$ and $\eta/m_{\rm Pl}$, on the
possibility of inflation starting from an initial monopole field
configuration in equilibrium without
gravity.  It is worth repeating that the numerical studies
\cite{lee1,breit,lee2}  already give significant constraints, as of course
does the work of \cite{sak1}.
\newline
\smallskip

\section
{\bf Relation of $\lambda/e^2$ to negative
gravitational mass density}

Let us now examine the question of possible monopole core inflation
for specified $\lambda/e^2$, assuming the
solution in the absence of gravity as an initial state, with
gravity then added.  A prerequisite to produce inflation, i.e., an
exponential growth with time in the volume of some region of space,
is the presence of negative gravitational mass density.  For a
Friedmann-Robertson-Walker spacetime, the driving force for
inflation is well known to be a negative value for the combination
$\mu=\rho + 3p$, where $\rho$ is the energy density, $p$ is the
pressure, and this expression defines
$\mu$ as the gravitational mass density. For such an isotropic
system this is equivalent to
\begin{equation}
\mu = T_{00}+\sum_{i=1}^3 T_{ii} \ \ , \label{gm}
\end{equation} where $T_{\mu\nu}$ represents the non-gravitational
(i.e., matter) contributions to the energy-momentum tensor. Sakai
\cite{sak1} emphasizes the form (\ref{gm}) for
$\mu$ and utilizes it as a way of determining when and where
inflation (or collapse) should be expected.  Indeed, (\ref{gm})
applies quite widely; for example, it holds in the case of a photon
gas in the vicinity of a large mass such as the sun or the earth
\cite{asg'}.  It is useful for our problem, because except at the
exact center of the monopole spatial isotropy is lost.

We want to determine the conditions for $\mu$ to be negative, so
that self-gravitational effects might cause inflation in the
monopole core. Consider the same scalar and gauge fields as above,
only omitting the gravitational field. The energy-momentum tensor
of the system is:
$$T_{\mu \nu} = D_{\mu} \Phi^a D_{\nu}
\Phi^a - g_{\mu \nu}
\frac{1}{2} D_{\sigma} \Phi^a D^{\sigma} \Phi^a -
\frac{1}{4} g_{\mu \nu} F^2$$
\begin{equation}  - F_{\mu \rho} F^{\rho}_{\,\,\,\, \nu}
- g_{\mu \nu} V(\Phi) \ \ , \label{emt}
\end{equation} where $g_{\mu \nu}={\rm diag}(-,+,+,+)$. As we are
interested in the parameter range
$\frac{\lambda}{e^2}\ll1$, let us consider as a starting point
 the static solution in the BPS limit
$\frac{\lambda}{e^2}\rightarrow0$. One can easily check, using the
BPS equation
$F^a_{ij}=\epsilon_{ijk}D_k\Phi^a$ (the indices
$i,j,k$ run only over the spatial components), that the sum of the
principal pressures vanishes. Hence the gravitational mass density
$\mu = \rho +
\Sigma T_{ii}$ is strictly positive. Therefore, in this limiting
case inflation is not possible. In fact one may notice that in the
BPS limit every component of the space-space part of the stress
tensor is zero:

\begin{equation} T_{ij} = \frac{\delta S}{\delta g_{ij}} = 0  \, \,
\,
\, \, \,
\,
\, \, \forall i,j=1,2,3 \ \ , \label{space}
\end{equation} a consequence of the topological nature of the
solution in this case (metric-independent form of the energy).
Furthermore, as the solution is static in the gauge $A_0 = 0$, we
have
$\,T_{0i} = T_{i0} = 0$ $\, \,
\forall i = 1,2,3$, which, along with (\ref{space}), implies that
the only nonvanishing component of the energy-momentum tensor is
the energy density
$T_{00}$.
\newline

To study the case of nonzero $\lambda$, we need some inequalities
on two contributions, from the gauge field and from the scalar
field.  As we have
\begin{equation}
\mu = \frac{1}{2} F^2 - 2V,  \label{FV}
\end{equation} to show that
$\mu$ is everywhere greater than zero we require a function which
is a lower bound on the gauge field contribution $F^2/2$, and a
function which is an upper bound on the scalar field potential $V$.   In the BPS limit the gauge field contribution
$F^2/2$ for the density in terms of the dimensionless Cartesian
coordinates
$\vec{x}=e\eta \vec{r}$ may be written
\begin{eqnarray}&& F^2/2=\frac{\eta}{ex^2}\frac{d}{dx}[({\rm coth}
\ x -
\frac{1}{x})(1-(\frac {x}{{\rm sinh}
\ x})^2)] =\frac{\eta}{ex^4}\times
 \nonumber  \\ \nonumber \\ &&
[(1-(x/{\rm sinh} \ x))^2 +2(x{\rm coth}
\ x-1)^2(\frac{x}{{\rm sinh} \ x})^2)] \ \ .
\end{eqnarray}  A simple function which never is greater than this
(but equal for
$x=0$, and discrepant only by ${\cal O}(1/x^6)$  for
$x\to\infty$) is
\begin{equation}  (F^2/2)_{\rm inf}=\frac{\eta}{e(3+x^2)(1+x^2)} \
\ .
\label{F}
\end{equation} Note that this expression is valid for
$\lambda=0$.  For nonzero
$\lambda$, one expects a compression of the monopole core.
Consequently, any change in (\ref{F}) will only be an increase, so
we indeed have a robust lower bound.

Now turn to the potential term.  In the vicinity of $x=0$, we have
\begin{equation}  -2V \ge
\frac{-\eta}{2e}(\frac{\lambda}{e^2})  \ \ ,
\end{equation} with the value decreasing monotonically in absolute
magnitude as
$x$ increases. This gives at $x=0$ the condition for
$\mu\le 0$
\begin{equation}
\frac{\lambda}{e^2}\ge 2/3 \ \ ,
\end{equation} as noted already by Sakai \cite{sak1}. Because at
this point
$2V$ attains its maximum, one might think that this is an absolute
condition.  However, it is easy to see that for the BPS ansatz
$2V$ falls more slowly with $r$ than $F^2$, opening the
counterintuitive possibility that $\mu$ may be negative away from
the center of the monopole for an even smaller value of
$\lambda/e^2$.  To address this question, we need to
obtain an improved bound on
$-2V$ away from the origin (where it is fixed by the topological
condition $\Phi(0) = 0$).

At large $x$, with the BPS solution substituted into $V$, we have
\begin{equation} -2V \approx
\frac{-2\eta}{e}(\frac{\lambda}{e^2x^2})  \ \ .
\end{equation} This is the form used as a starting ansatz by
Sakai.  However, for nonzero
$\lambda$ small fluctuations about the vacuum have a mass
$m=\sqrt{2\lambda}\eta$.  A conservative lower bound on the
asymptotic behavior therefore is
\begin{equation} -2V_{bound} \approx
\frac{-2\eta}{e}(\frac{\lambda}{e^2x^2})
e^{-2\sqrt{2(\lambda/e^2)}x }\
\ .
\end{equation} This Yukawa falloff actually gives an upper bound
on the magnitude of $V$ because the effective source at the
geometric center of the Yukawa field
$\Phi(x)$ is weakened by the same quartic potential which gives
mass to the long-distance small fluctuations.  The exponential
falloff is important in principle, because otherwise at
sufficiently long distances $-2V$ clearly would overcome $F^2/2$.
Let us combine the expressions for small and large $x$ to get an
infimum for $-2V$. Let us take at each $x$ the less negative of our two
forms, matching them at the crossover point, which for infinitesimal
$\lambda$ occurs at
$x=2$:
\begin{eqnarray}&& -2V_{\rm inf} =
\frac{-\eta}{2e}(\frac{\lambda}{e^2})
\ ,  \ \ x\le 2 \nonumber
\\  \nonumber \\ && =\frac{-2\eta}{e}(\frac{\lambda}{e^2x^2})
e^{-2\sqrt{2(\frac{\lambda}{e^2})}x} \ , \  \ x>2 \ \ .
\end{eqnarray} For the interval $x\le 2$ the strongest condition
for positivity of $\mu$ comes at the  end ($x=2$):
\begin{equation}
\frac{\lambda}{e^2}\le 2/35 \ \ .
\end{equation} At larger $x$, we may simplify initially by using the
asymptotic form for $F^2/2$, yielding the requirement
\begin{equation}
\sqrt{2\frac{\lambda}{e^2}}xe^{-\sqrt{2\frac{\lambda}{e^2}}x}\le 1
\ \ .
\end{equation} As the maximum of this expression is $1/e$ for
$\sqrt{2\frac{\lambda}{e^2}}x=1$ or
$x\approx 3$ with the previously obtained value for
$\lambda/e^2$, and at that value of $x$ the infimum of
$F^2/2$ is smaller than the asymptotic form by a factor
$81/120 \approx .68 > 1/e= .36$, we see that with ample assurance,
for
\begin{equation}
\frac{\lambda}{e^2}\le 2/35\approx .06  \ \ ,
\end{equation} the gravitational mass density $\mu$ of the 't
Hooft-Polyakov monopole is nowhere negative, so that the most
elementary criterion for inflation is not satisfied.

In particular, if gravitational coupling were to be introduced at
this point, and then increased, it could produce compression, even
collapse, but never inflation. While we have a rigorous lower
bound on the critical value of $\lambda/e^2$, it would be helpful
to increase the bound.  The following semi-quantitative argument
assures that the actual bound must be larger by an order of
magnitude.  For the BPS solution, the gauge field energy and the
gradient energy of the Higgs field are identically distributed,
but as $\lambda/e^2$ increases the Higgs gradient energy is
suppressed in the external region.  Consequently, a $1/R$ term in
the total energy, where $R$ is the radius below which $F^2$ stops
increasing as $1/r^4$, has its coefficient reduced by as much as
$\frac{1}{2}$, while the coefficient of a term proportional to
$R$, associated with the integral of the gradient energy inside
$R$, is roughly doubled.  This means that the equilibrium $R$ is
reduced by a factor of about two, so that the central density
$F^2/2$ of the gauge field contribution to $\mu$ is increased by
at least an order of magnitude.   When this is inserted into the
previous estimates, it leads immediately to the claimed order of
magnitude increase in the critical value of $\lambda/e^2$, which
therefore surely is at least $0.5$ - the same order of magnitude
as the smallest values considered in \cite{sak1}. This claim could
be tested numerically if Sakai's calculation were repeated using
as a starting point the static solution to the Yang-Mills-Higgs
equations including the $\lambda$ term (a solution which itself is
determined numerically \cite {'t H}).
\newline
\smallskip

\section
{\bf Influence of $\eta$ on the possibility of inflation}

Let 
us  turn now to consider the other (dimensionful) parameter in
(\ref{sakai}). The fact that the vev
$\eta$ of the scalar field must be at least of the order of the
Planck mass  to have inflation is commonly accepted. It was
identified in \cite{lin2} and \cite{vil2}, for example, as required
by the slow-roll condition for new inflation ($|\ddot{\Phi}|
\,
\ll 3H\dot{\Phi}$ and
$\dot{\Phi}^2 \ll V(\Phi)$). However, the usual inflationary
considerations assume that the scalar field is constant throughout
the whole space, or at least within some big region. Here is a simple proof that
$\eta \approx m_{\rm Pl}$ is a  necessary condition for inflation in
the case under consideration (in which the negative gravitational
mass density is restricted to a region in the core of the
monopole), based on a paper by Blau, Guendelman, and Guth
\cite{blau}.  Because from the result above negative $\mu$ implies
$\frac{\lambda}{e^2}\approx 1$, in the following let us assume that
this condition is met.

In \cite{blau} the authors consider a spherically symmetric bubble
of false vacuum separated by a domain wall from an infinite region
of true vacuum. They obtain a critical mass $M_{\rm cr}$ such that
for mass $M$ of the bubble bigger than
$M_{\rm cr}$ the bubble will inflate, while for $M < M_{\rm cr}$ it
will collapse. If one neglects the surface energy density on the
bubble wall, then from (5.14) in
\cite{blau} one sees that the critical mass is
\begin{equation} M_{cr} = \frac{4 \pi}{3} \chi^{-3}
\rho,
\end{equation} \\ where $\chi$ (in the notation of \cite{blau}) is
the Hubble parameter of the de Sitter region inside the bubble and
$\rho$ is the energy density of the false vacuum. This simply means
that the radius of the false vacuum region should be bigger than
the distance to the horizon in the de Sitter space for inflation to
be possible. Let us now consider the core of the 't Hooft-Polyakov
monopole as a spherical region of false vacuum with average energy
density
$\bar{\rho}$. As the mass of the monopole is $M =
\frac{4 \pi \eta}{e} f(\frac{\lambda}{e^2})$, where
$f(\frac{\lambda}{e^2})$ depends very little on its argument and is
$f(\frac{\lambda}{e^2}) \approx 1$, we get for the average energy
density:
\begin{equation}
\bar{\rho} = 3 e^2 \eta^4, \label{rho} \ \ ,
\end{equation} using the fact that the equilibrium
radius of the monopole is $R = 1 / e
\eta$. As the Hubble parameter H is given by
\begin{equation} H^2 = \frac{8 \pi G \bar{\rho}}{3} \ \ ,
\end{equation}   from the condition for inflation $R > H^{-1}$
follows:
\begin{equation}
\eta > \sqrt{\frac{1}{8 \pi G}} \ \  ,
\end{equation} using (\ref{rho}). In units
$\hbar = c = 1$:
\begin{equation}
\eta > \sqrt{\frac{1}{8 \pi}} m_{\rm Pl} \approx 0.2 m_{\rm Pl}.
\end{equation} This rough estimate agrees pretty well with the
numerical calculations in \cite{sak1}, which give
$\eta > 0.3 m_{\rm Pl}$.  The idea that a topological defect can
inflate if its size is bigger than the cosmological horizon is
suggested already in \cite{vil2}, but without derivation.

The above estimate of the critical value of $\eta$ would be accurate if as
in \cite{blau} the entire mass of the bubble were associated with the
Higgs self-coupling, but that is not true here because of the positive
gauge field contributions to $\mu$, which surely will increase the
threshold further. Again, this could be checked by numerical
work with the flat-space solution for nonzero $\lambda$ as a
starting point.
\newline
\smallskip

\section
{\bf Influence of primordial topological zeros on inflation}

We have seen that starting out with a monopole in equilibrium and
then adding gravity will not yield inflation unless strict
conditions are satisfied. However, as gravity presumably is always
present, we should consider the early stage of a process which
would have generated monopoles in the absence of gravity, but now
with gravity included.  Let us imagine an initial condition introduced
at some stage in which the Higgs field everywhere has a very small
expectation value, but the minimum of the Higgs self-coupling
potential, i.e., the equilibrium vev, has a substantial nonzero
value.  From the continuity of the small yet fluctuating
expectation value, we expect topological zeros.  The canonical view
\cite{lin2,vil2} is that because this gives an initial condition
for inflation, the gradient of the Higgs field in the neighborhood
of each zero will rapidly decrease in magnitude, so that the field
will be pinned near zero indefinitely by the topology, thus
generating eternal inflation.

This issue has been addressed in numerical calculations for models
in 2+1  dimensions by Linde and Linde [LL] \cite{lin2}, using quite
large values for the parameters, so that one would be at least near the
classical threshold for inflation of a soliton (vortex in this case),
initially in equilibrium without gravity, when gravitational coupling is
introduced.  Even then, when counting inflationary zones associated
with maxima of the Higgs potential, LL found only a minority of those
zones carried topological zeros, so that even under these optimal
circumstances  (large $\lambda$)  topological zeros at most make
a minor quantitative enhancement in the rate of inflation.

Let us approach the question here in a different way, which
should be complementary but may serve to highlight the
circumstances in which topological inflation does or does not take
place.  The process which eventually can produce normal vacuum
invokes quantum fluctuations to accomplish this task.  This
implies that for consistency we should treat quantum-mechanically
all degrees of freedom of the Higgs field configuration, including
the spatial coordinates of each topological zero.  The same
inflationary effect which flattens the gradients of the field
treated classically also spreads the probability distribution
associated with what classically would be a localized zero.

Consequently, even though the zero exists, it becomes possible
that it has little influence on the fluctuation process which can
lead to regions of normal vacuum.  When that is correct, then
sooner or later a zero might find itself in such a normal region,
clothed with the gauge field configuration required to make its
total mass finite. In that case the considerations above become
relevant, because a monopole in approximate equilibrium will only
be able to inflate if the values of the parameters $\lambda/e^2$
and $\eta$ are appropriately large. Thus, the zero will be
eternal, but inflation in the vicinity of the monopole need not
be.  One may argue that this reasoning is robust, because classical
physics always is an approximation to quantum physics.  If
treating a degree of freedom as quantum-mechanical gives different
results from treating it classically, that immediately implies the
classical approximation is invalid, and the quantum description is
necessary.

The earlier discussions in this paper appear not to be found in
the literature, but nevertheless are quite straightforward.  On the
other hand the argument just given about quantum behavior of
topological zero coordinates may be less obvious.  Here are some
questions it raises: In the standard approach leading to the
appearance and significance of topological zeros \cite{kib}, it is
quantum fluctuations which produce the zeros.  Indeed (as pointed
out in \cite{lin2,vil2}), continuing quantum fluctuations during
the epoch of very small Higgs field will be generating additional
zeros.  Further, for a zero at the center of a soliton in
equilibrium -- an object which is well-described by a stable,
classical field configuration -- quantum fluctuations of the
position are unimportant because the Compton wavelength of the
soliton is small compared to the internal dimensions of the
object.  Thus one could ask both ``Don't we already have a fully
quantum description?" and ``Isn't the location of the zero anyway
a classical degree of freedom?"

To address both questions, we need to look at the early evolution
of a topological zero -- before it has `nucleated' a stable
configuration.  Of course, if the couplings are such that a
 monopole stable without gravity would inflate with gravity taken into account,
 then there isn't a stable configuration,
but we now are investigating the opposite case, so that only early
development is possibly relevant.  It is obvious that the zero
appears during a period when inflation is occurring, precisely
because the necessary fluctuations depend on the Higgs field being
small, and therefore close to a maximum in the Higgs potential.
Thus the issue is whether the presence of topological zeros will keep inflation going
longer than it otherwise would, e.g., in a theory with a single
Higgs field constrained to be nonnegative, with maximum potential at zero value of that field.

To simplify the problem take an Ansatz for the Higgs field of the
form
$$\Phi(\vec{r},t)=\Phi_0\sin(|\vec{r}-\vec{r_0(t)}|/\ell),
|\vec{r}-\vec{r_0}|\le\ell\pi /2$$
\begin{equation}
\Phi(\vec{r},t)=\Phi_0, |\vec{r}-\vec{r_0}|\ge\ell\pi /2 \ \ .
\label{phi0}
\end{equation}

In the light of the comment about the classical character of the
zero location for a stable soliton, to address this issue we need
to examine the behavior of the coordinate for a zero in the
presence of a very small Higgs field.  Let us do this initially
while neglecting gravity, adopting the following method:  Assume
the Higgs field profile has the same shape and size it would have
in the static limit, except that the magnitude of the field at
asymptotic distance from the zero is much smaller.  Under rigid
motion of this structure (neglecting the gauge field contribution
which presumably only is important for large Higgs field and
stable monopole), what is the kinetic energy, and hence the
inertial mass corresponding to motion of
 the coordinate $\vec{r}_0$? This quantity is evaluated by substituting
 into
the energy density expression given by (\ref{emt}) the form (\ref{phi0}),
and evaluating the integrated density as $m_0(d\vec{r}_0/dt)^2/2$:
\begin{equation}
m_0=\frac{\pi^2}{6}\bigg{(}\frac{\pi^2}{6}-1\bigg{)}
\Phi_0^2\ell  \ \ .
\end{equation}
Evidently for $\Phi_0\ell \ll 1$ this gives rise to an uncertainty
in position of the zero much greater than the size $\ell$ of the
region in which the field departs from its temporarily small
`vacuum' value.  Even if one made the extreme assumption that
there were only one scale in the problem, given by the Hubble
length, then still the uncertainty would be at least comparable
with the size of the causally connected region.

A consequence of this proposed uncertainty in position is that the
expectation value of $\Phi^2(x)$ will be nearly constant, and an
appreciable fraction of $\Phi_0^2$ throughout the region, rather
than vanishing at some point.  Thus, if this view were correct, the
evolution of $\langle\Phi^2(x)\rangle$ would be quite similar to
that for a case where there was no zero present in the particular
region.  This seems an attractive picture, suggesting that there should be very little difference between behavior
near a local maximum of the potential with, or without, a topological zero.  Topological zeros clearly have global definition, but
nevertheless correspond to small, localized disturbances.  If so,
they should be able to produce inflation in their cores if and
only if a monopole in equilibrium without gravity would be at
least close to the threshold for generating core inflation.  The
creation of these zeros should not significantly enhance inflation
during the early evolution of the inflaton field away from the
maximum potential at zero field.

 An interesting numerical test of the importance of topological inflation could
 be obtained by repeating the numerical work in \cite{lin2} with an artificial
 requirement that the value of the complex Higgs field lie in the upper half of the complex
 plane, and that if in its evolution the field at some point would go into the lower
 half plane then the value at the next step is taken as the reflection of the forbidden
 value back into the upper half plane, with the time derivative of the imaginary 
 component also reversed in sign.  Evidently there is no topological charge for
 such configurations, but it might be that inflation would proceed just as effectively
 in this case.  This would mean that topology would be essentially an accidental
 attribute of certain configurations involving small values of the Higgs field which generate inflation.  
 
 In fairness to LL, a comment seems in order.  Their random process for generating values of the complex field in principle should capture all kinds of quantum fluctuation, therefore including quantum fluctuations in positions of zeros in the field.
 Thus, one should already be able to see what is claimed here, the suppression of influence by the zeros on inflation, in their calculation.   That indeed appears to be the case, because most of the local maxima in the Higgs potential are associated with near zero values of the field, but not topological zeros.

How would a global monopole, which according to the earlier
discussion could inflate, acquire the gauge-field structure which
for phenomenologically indicated parameters would preclude
inflation?  If seems possible that there would be competition between
the launching of inflation and the development of the gauge-field
structure.  Near the center of the monopole, and even at large
radii, spontaneous gauge-field fluctuations which lower gradient
energy of the scalar field should gain strength. The inertia of
such a fluctuation in a spherical shell of fixed radius is
proportional to the thickness of the shell, and inversely
proportional to the gauge coupling $e^2$, while the `driving
potential' -- the scalar field gradient energy -- also is
proportional to the thickness. Thus this process should proceed at
a rate reduced from the maximum allowable only because of the
small magnitude of $e^2$, and still should be enormously fast
compared to the inflation rate for the phenomenologically
indicated much smaller value of $\lambda$.

\smallskip

\section
{\bf  Role of supersymmetry}

It is well-known that in inflationary models based on
supersymmetric (susy) extensions of the Standard Model the quartic
self-coupling of the inflaton field can be strongly suppressed.
(For an extensive recent review of susy inflationary models the
reader is referred to \cite{lyth}.) With exact global
supersymmetry, the potential is typically independent of some of
the fields i.e., it has flat directions in field space. Due to the
slow-roll condition for inflation, those fields are exactly the
candidates for the inflaton field, and because of the flatness of
the potential one has
$\lambda = 0$, independently of whether supersymmetry is broken or
unbroken and if broken whether spontaneously or softly.  If susy is present
in nature, it is expected to be local so as to accommodate gravity. 
That is, supersymmetry and gravity only can be reconciled in the form of supergravity.  It is known that in supergravity the flatness of global susy is lifted:
$\lambda$ is of the order of $M_s^4/ m_{\rm Pl}^4$, where
$M_s$ is the supersymmetry breaking scale. If $M_s \ll m_{\rm Pl}$, then
$\lambda$ is strongly suppressed, though not identically zero.
Hence in global or local susy inflation of the monopole core must depend on a different type of effective potential
for the Higgs field from the `conventional' one described here.  The considerations above do not apply
directly, but one would expect them to give a correct qualitative
analysis of the phenomenon, most probably meaning no inflation.

 Part of this qualitative reasoning is the following:  In exact
susy, one expects the BPS bound on the mass to be saturated, hence
no inflation of the monopole core, indeed, no inflation at all.
If there is some breaking, then there will be some potential for
the Higgs field other than the $\phi^4$ term. In a homogeneous
space this would be enough for inflation. However in the nontrivial
background provided by an 't Hooft-Polyakov monopole the
phenomenological requirement that the couplings in this potential
be small for compatibility with CMB data \cite{lyth} implies that
the $F^2$ term in (\ref{FV}) would dominate over the potential
term, and hence inflation of the monopole core would be forbidden.
\newline
\smallskip

\section
{\bf Conclusions}

We have seen that for observationally indicated values of
$\lambda/e^2$ the gravitational mass density $\mu$ of an 't
Hooft-Polyakov monopole is everywhere positive, so that without
some new factor there could not be monopole inflation.  As
mentioned, introducing gravity would only strengthen the result by
spatially compressing the system, further increasing $\mu$, and
eventually leading towards a Reissner-Nordstrom black hole, rather
than inflation (again modulo the solutions of \cite{breit} which
have no flat space-time limit). Having confirmed a simple-minded
and intuitive expectation, we still should explore possible ways
of resuscitating the notion.  A first possibility is quantum
fluctuations leading to an expanded monopole in which the gauge
field energy is diluted.  If indeed $\lambda/e^2$ is only somewhat
smaller than unity, such fluctuations surely become plausible.  A
second possibility is two-stage inflation, where the first stage
involves the Higgs field as inflaton field (with large $\lambda$, 
meaning no supersymmetry), possibly producing
monopole core inflation, while the second involves a
separate, non-gauge-coupled inflaton field with small $\lambda$
\cite{lin2}. Clearly, this would make our connection to the core
region of the monopole exceedingly indirect.   It would mean that
the core of the monopole would be exponentially large compared
the size of our visible universe.

In summary,  for a monopole in equilibrium without gravity, the
slow or fast introduction of gravitational couplings could not
precipitate inflation through the classical equations of motion
unless the parameters were above the threshold values
$(\lambda/e^2)_{\rm thresh}
\simeq 1$ and $\eta_{\rm \ thresh}
\simeq m_{\rm Pl}$. These values are a bit higher than suggested
by Sakai \cite{sak1}, and a numerical test has been proposed to
verify the higher values.   For parameters only a bit below the threshold
values, quantum fluctuations still could initiate inflation.  The
singularity theorems mentioned earlier imply that, regardless of
the values of the parameters, quantum fluctuations are essential to
monopole core inflation.  The main result of the classical
computations is that not even quantum fluctuations could produce
such a phenomenon unless the parameters $\lambda$ and $\eta$ both were
sufficiently large.

For the starting conditions associated with a topological zero of
a field which initially has very small magnitude everywhere in some region, it has been argued
that quantizing the degree of freedom associated with the position
of the zero implies effective decoupling of the zero from the
spatial evolution.  If so, by the time the zero is treatable as a
classical coordinate, it likely will be the center of a monopole
field configuration moderately close to equilibrium, and so could
not inflate unless the model parameters were at least close to the
threshold values.  
If they were close, then the eternal inflation
associated with the zero would be in a sawtooth pattern, where
inflation would start, the zero would eventually emerge in a
noninflating region, inflation would reignite producing more
topological zeros, and so on.

 At first sight, the claim about the 
influence of quantum uncertainty sounds quite different from earlier
work, including the numerical lattice studies of quantum fluctuations
by LL \cite{lin2}, but a more helpful view of these very
low-mass topological defect `embryos' may be to say that not only their
positions but also their numbers show large quantum fluctuations, as 
found in \cite{lin2}.  Thus the position uncertainty may be viewed as only one aspect
of the rapid pair creation.  This has interesting implications, including
the ability of net topological charge to shift position faster than
the speed of light.  It makes a problem of which comes first:  Small
magnitude of $\Phi_0$ allowing copious production of topological zeros,
or topological zeros plus inflation enforcing small $\Phi_0$ and hence
continued inflation.  Based on the quantum uncertainties, one can argue
for the former as the dynamically significant statement.  The results of
LL quoted earlier indicate that the topological zeros at most give a 
modest quantitative enhancement to the maintenance of inflation.  Even
those results were for a large value of the self-coupling $\lambda$, where
as noted earlier one already would expect  a monopole in equilibrium to inflate once coupling to gravity is included.  A numerical test has been proposed to check whether, even for the configurations with topological zeros,  the topology is more than an accidental aspect of the fact that the Higgs field comes close to a local maximum in the Higgs potential.
Boubekeur and Lyth  \cite{Boubekeur:2005zm}  have argued that almost all viable models of inflation involve starting from a `hilltop', that is, the near neighborhood of a rather flat maximum in the inflaton potential.  For this purpose it shouldn't matter whether at some point the potential actually achieved its maximum, which for a locally analytic solution would  mean a topological zero.

If supersymmetry were broken only at
low scales, then monopole inflation from equilibrium would be strongly
suppressed, and by the above argument a topological zero of the
Higgs field no longer should be expected to generate inflation.
Thus, if this understanding of the implications of quantum physics
for zeros of a Higgs field is correct, and if phenomenologically
indicated parameters of inflationary cosmology should be accepted
at face value, then monopole topological inflation, while
conceptually instructive and appealing, does not provide a
compelling alternative or even a significant enhancement to
chaotic eternal inflation \cite{lin1,steinh1,vil1}.  However,
domain wall inflation in a theory with a single inflaton field
remains a viable and important mechanism for eternal inflation.
In this case is  there is no
static-gradient positive contribution to the gravitational mass density to
compensate for the negative density coming from the scalar field
potential.  Further,   if the domain wall achieves large area, it also achieves
large inertial mass,, so that the argument about quantum uncertainty in its position disappears. By the same token it becomes harder to envision copious pair creation of walls and antiwalls.   
 
\smallskip

The idea for this work came from a discussion with Lilia Anguelova, who
wrote an initial draft including many of the 
essential calculations.  She has repeatedly provided instructive 
reactions to successive drafts, and deserves much credit for any
merit of the final result.
I thank Martin Bucher and Martin Ro\v{c}ek for useful comments,
 and Alexander Vilenkin for penetrating reactions
to an early draft.  Andrei Linde generously and aptly criticized
not only that early draft but also repeated revisions.
This work was supported in part by the National
Science Foundation, Grant PHY-0140192.

\bigskip
\end{document}